\documentclass[a4paper,fleqn]{cas-sc}

\usepackage[numbers]{natbib}

\usepackage{physics}
\usepackage{siunitx}
\usepackage{scalerel}

\def\tsc#1{\csdef{#1}{\textsc{\lowercase{#1}}\xspace}}
\tsc{WGM}
\tsc{QE}

\begin{document}
\let\WriteBookmarks\relax
\def\floatpagepagefraction{1}
\def\textpagefraction{.001}

\shorttitle{LSTMs for Anomaly Detection in Magnet Power Supplies}    

\shortauthors{I. Lobach, M. Borland}  

\title [mode = title]{Long Short-Term Memory Networks for Anomaly Detection in Magnet Power Supplies of Particle Accelerators}  

\author{Ihar Lobach}[orcid=0000-0001-9724-6880]
\cormark[1]
\cortext[1]{Corresponding author}
\ead{ilobach@uchicago.edu}

\author{Michael Borland}[orcid=0000-0003-2536-3476]

\affiliation{organization={Advanced Photon Source, Argonne National Laboratory},
            addressline={9700 S Cass Ave}, 
            city={Lemont},
            postcode={60439}, 
            state={IL},
            country={USA}}

\ead{mborland@anl.gov}




\begin{abstract}
This research introduces a novel anomaly detection method designed to enhance the operational reliability of particle accelerators---complex machines that accelerate elementary particles to high speeds for various scientific applications. Our approach utilizes a Long Short-Term Memory (LSTM) neural network to predict the temperature of key components within the magnet power supplies (PSs) of these accelerators, such as heatsinks, capacitors, and resistors, based on the electrical current flowing through the PS. Anomalies are declared when there is a significant discrepancy between the LSTM-predicted temperatures and actual observations.
Leveraging a custom-built test stand, we conducted comprehensive performance comparisons with a less sophisticated method, while also fine-tuning hyperparameters of both methods. This process not only optimized the LSTM model but also unequivocally demonstrated the superior efficacy of this new proposed method. The dedicated test stand also facilitated exploratory work on more advanced strategies for monitoring interior PS temperatures using infrared cameras. A proof-of-concept example is provided.
\end{abstract}


\begin{highlights}
\item Presents a novel LSTM-based anomaly detection method for enhancing particle accelerator reliability by monitoring magnet power supply temperatures.
\item Demonstrates the proposed method's superior efficacy through comprehensive performance comparisons.
\item Introduces advanced temperature monitoring using infrared cameras, providing a proof-of-concept example.
\end{highlights}


\maketitle

\section{Introduction}\label{sec:intro}

Particle accelerators are essential tools in both research and industrial applications, coming in different shapes and a variety of sizes ranging from a few meters to tens of kilometers. Regardless of their design, they all utilize magnetic fields to steer and focus charged particles, which propagate as a beam. These magnetic fields are typically generated by electromagnets, as opposed to permanent magnets.
Each electromagnet (or a small cluster) is powered by a dedicated power supply (PS), which transforms AC wall power into the DC current required by electromagnets for operation. Large particle accelerators can require thousands of such power supplies. The malfunction of even a single PS can lead to a complete loss of the particle beam, and, consequently, to costly downtime for the entire facility, highlighting the importance of reliability of the power supplies.

The Advanced Photon Source (APS) \cite{galayda:pac95-mad02}, a large-scale electron storage ring with an approximate circumference of 1.1 kilometers, generates intense X-ray beams, empowering advanced research in materials science, chemistry, and biology through precise atomic and molecular level investigations.
It was in operation from 1996 to 2023.
On average, there were about seven downtimes per year caused by power supply failures \cite{aps-run-history}. They were up to 30 hours long with an average of about 2 hours.
Based on the yearly budget and the number of scheduled operating hours, each hour of APS operation was worth approximately 30 thousand US dollars; the actual value is significantly greater, since the APS supported more than 50 simultaneous experiments.
An anomaly detection method, capable of detecting precursors for complete power supply failures, would significantly improve reliability of the storage ring by giving the PS technicians an opportunity to repair the PS before it fails. It would help avoid undesirable interruptions of APS users' experiments and costly downtimes.
The APS storage ring underwent a major upgrade in 2023--2024, with the new ring being commissioned at this time.
The magnets and the power supplies of the storage ring are receiving substantial enhancements. This upgrade presents an opportunity to revisit and innovate anomaly detection methodologies for the power supplies of APS, leveraging the upgraded systems' enhanced monitoring capabilities and capitalizing on the significant strides made in anomaly detection methods over the last few decades.
The main goal of this paper is to propose an effective and robust anomaly detection method for one or several parameters of the power supplies, which could be scaled to the thousands of the power supplies of the upgraded APS storage ring (APS-U) in the near future. A secondary goal of this paper is to present a proof-of-concept example for a long-term vision of implementing more advanced anomaly detection techniques (Section~\ref{sec:ir_camera}).

\section{Background}
\begin{figure} 
	\centering
		\includegraphics[width=\textwidth]{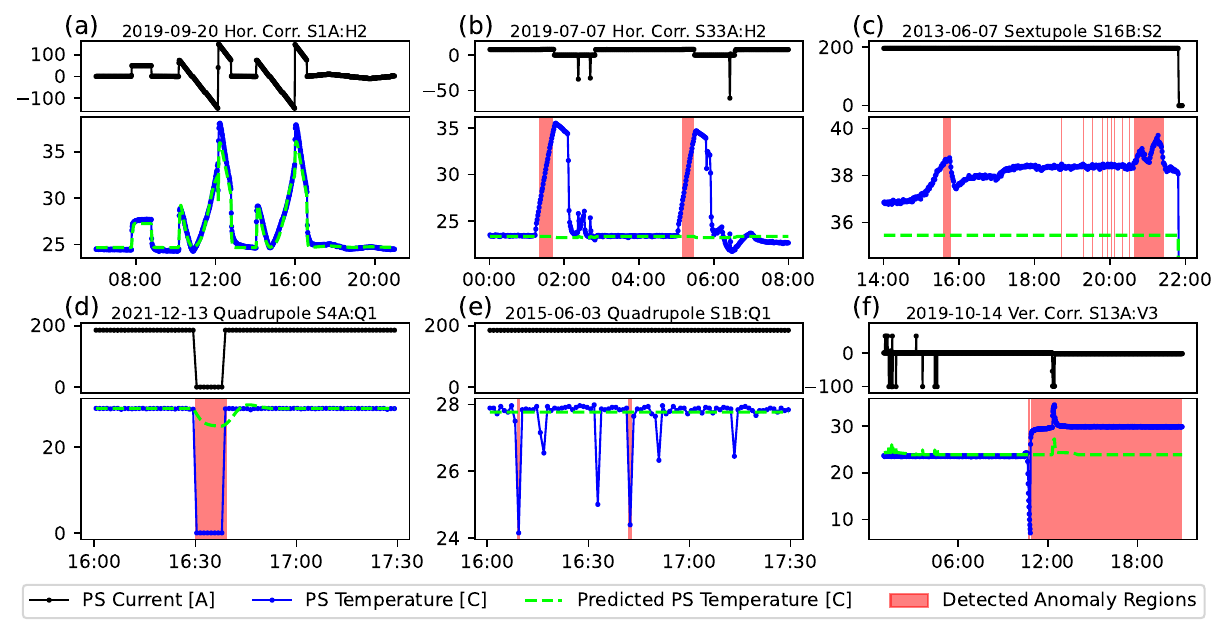}
	  \caption{Examples of anomalies detected with the
      LSTM-based method in the historical APS data from 2008 to 2023.}\label{fig:aps_historical_data_examples}
\end{figure}
As the first step, we analyzed 15 years of historical data of APS operation \cite{aps-run-history,lobach:napac2022-tupa29,lobach:ipac2023-thpl005} for 1320 storage-ring power supplies \cite{aps-magnets,decker:pac95-far13}. This investigation provided valuable insights into (1) the types of anomalies likely to occur in APS-U power supplies, (2) critical power supply parameters that require monitoring, and (3) the anomaly detection method that holds the most promise.
Our findings indicate that anomalies in PS temperature parameters are the most reliable signs of PS malfunction, whereas trying to detect abnormalities in PS currents or voltages results in large numbers of false positives. According to our analysis, the most promising method for anomaly detection in PS temperatures is the following.
Using the normal-operation data, we train an LSTM model that takes a time-series of the PS current as an input and returns a
time-series of the PS temperatures as a prediction (corresponding to the same timestamps as the input sequence). Then,
the predicted temperatures are compared with the measured ones; and an anomaly is declared if they deviate from each
other significantly. Several examples for this method are shown in Fig.~\ref{fig:aps_historical_data_examples}. Panel
(a) presents a part of the time interval used for training the model. The example in Fig.~\ref{fig:aps_historical_data_examples}(b) corresponds to a stuck mixing
valve for the cooling water (two incidents). When cooling water stopped
circulating, the temperatures of multiple PSs  increased quickly. When one
of them reached \SI{50}{C}, it tripped (by design).
Figure~\ref{fig:aps_historical_data_examples}(c) shows a peculiar anomaly in
one PS, which tripped shortly after.
The anomalies detected in Fig.~\ref{fig:aps_historical_data_examples}(d),(e), and (f) 
do not immediately lead to a PS trip. In
Fig.~\ref{fig:aps_historical_data_examples}(d), the temperature and current readings briefly went to
zero, which could be a network problem. In
Fig.~\ref{fig:aps_historical_data_examples}(e), the anomalies flag
a very noisy signal from the temperature sensor. Perhaps the contact between the sensor
and the surface
was poor. The anomaly type shown in
Fig.~\ref{fig:aps_historical_data_examples}(f) may be observed, for example, when a PS is replaced by a new one. The captions for the panels in Fig.~\ref{fig:aps_historical_data_examples} indicate the dates of the presented time intervals and the names of the magnets, powered by the power supplies under consideration.

In the subsequent sections of this paper, we will focus on the proposed LSTM-based approach for anomaly detection in PS temperatures with the intention of future application in the power supplies of the upgraded storage ring (APS-U).
Our analysis of historical APS data was limited by the data logging rate of one point every 64 seconds. Further, we were constrained by the existing data and could not compile datasets ideally suited for LSTM model training. 
For the APS-U power supplies, we plan to increase the data rate to \SI{1}{Hz} and choose more optimal training datasets, as detailed in Section~\ref{sec:psTestStand}.
Further, there will be improvements in the temperature sensor data quality and in the record-keeping of PS swap-outs. This will reduce the number of anomalies shown in Fig.~\ref{fig:aps_historical_data_examples}(d),(e),(f), which are still useful for scheduling preventive PS maintenance, but not as useful as the ones in Fig.~\ref{fig:aps_historical_data_examples}(b),(c), which require urgent actions to prevent an impending PS failure.
Our choices for the training dataset, architecture, and hyperparameters for the LSTM model within the context of APS-U, as well as performance benchmarking, will be presented across Sections~\ref{sec:psTestStand}-\ref{sec:benchmarking}.
Prior to this, we will provide a physics-based explanation of why LSTMs are powerful at predicting PS temperatures in Section~\ref{sec:theory}.

\section{Theoretical Foundations of the Proposed Anomaly Detection Method}\label{sec:theory}

The time evolution of the PS temperature $T$, which is also a function of the position inside the PS, can be
rigorously determined by solving the heat equation, 
\begin{equation}\label{eq:continuous_heat_equation}
    \rho c_p \pdv{T}{t} = q_V - \div{\left(k\grad{T}\right)},
\end{equation}
\noindent where $\rho$ is the mass density,  $c_p$ is the specific heat capacity, $q_V$ is the volumetric heat source, and $k$ is the thermal conductivity;
all of which are functions of the position inside the PS. In addition, $q_V$ is also a function of the PS current. It
describes the heat generated inside the PS due to its operation. Equation~\eqref{eq:continuous_heat_equation} must be
complemented by boundary conditions at the border of the PS, which will take into account the heat dissipation to the surroundings of the PS. It is important that only the solid
parts of the PS are considered with Eq.~\eqref{eq:continuous_heat_equation}, because it does not account for convection. There is some freedom in defining the border of the PS. One could  consider the board
and the components on the board as the PS, or one could also include the chassis (see Fig.~\ref{fig:PS_photo} in Section~\ref{sec:psTestStand}).

\begin{figure} 
	\centering
		\includegraphics[width=0.5\textwidth]{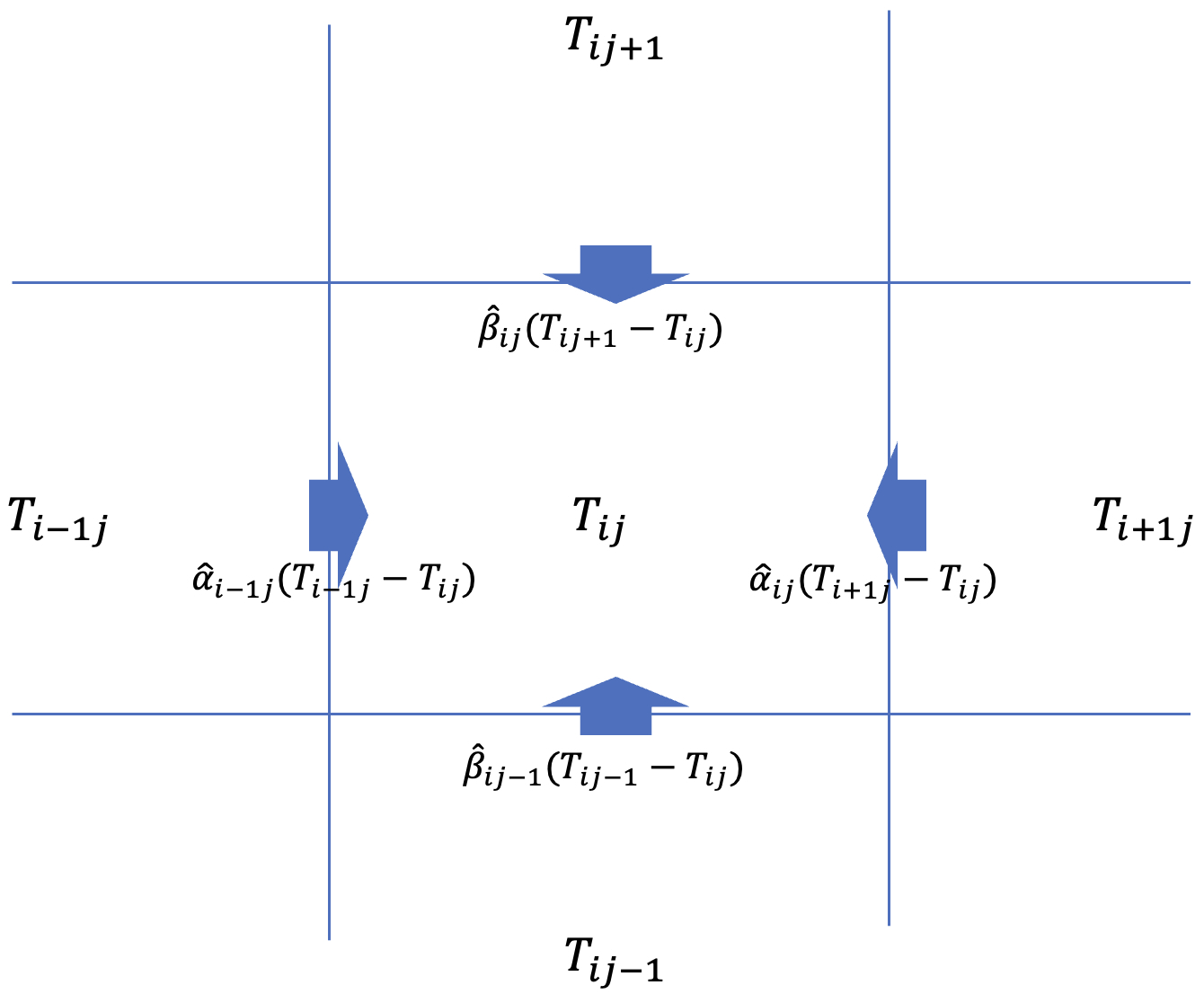}
	  \caption{Notation for numerical solution of the heat equation (example for two dimensions).}\label{fig:numerical_heat_equation}
\end{figure}

The heat equation can be numerically solved using the method of finite differences. It can be applied to a PS of any
shape, but here we will approximate it by a right parallelepiped and divide it into a large number of small cubes, for
the sake of simplicity. This is sufficient for delivering the main message of this section. The incremental change in the temperature from time $t$ to time $t+\Delta t$ of a cube with indexes
$i,j,k$ is
\begin{multline}
\begin{aligned}\label{eq:finite_difference_heat_equation_0}
C_{ijk} \Delta T_{ijk} = 
\Delta t\cdot\Bigl[P^{\mathrm{(heat)}}_{ijk}(I) - P^{\mathrm{(diss)}}_{ijk}(I,T_{ijk}-T_\mathrm{room})
&+ \alpha_{ijk} \; (T_{i+1\;j\;k} - T_{ijk}) + \alpha_{i-1\;j\;k} \; (T_{i-1\;j\;k} - T_{ijk})
\\
&+ \beta_{ijk} \; (T_{i\;j+1\;k} - T_{ijk}) +
\beta_{i\;j-1\;k} \; (T_{i\;j-1\;k} - T_{ijk})
\\
&+ \gamma_{ijk} \; (T_{i\;j\;k+1} - T_{ijk}) +
\gamma_{i\;j\;k-1} \; (T_{i\;j\;k-1} - T_{ijk})\Bigr]
\end{aligned}
\end{multline}
\noindent where $C_{ijk}$ is the cube's thermal capacity; $\alpha_{ijk},\,\beta_{ijk},\,\gamma_{ijk}$ are the coefficients for heat
 exchange between adjacent cubes. A diagram explaining the meaning of these coefficients for a two dimensional case is shown in Fig.~\ref{fig:numerical_heat_equation}. $P^{\mathrm{(heat)}}_{ijk}(I)$ is the heat generated inside the cube. $P^{\mathrm{(diss)}}_{ijk}(I,T_{ijk}-T_\mathrm{room})$ is the heat dissipated from the cube to the air, if this cube has a boundary(s) exposed to air. An assumption is made that the surface of the PS is in contact only with the surrounding air, and the power dissipated to the air is a function of the temperature difference $(T_{ijk}-T_\mathrm{room})$, where $T_\mathrm{room}$ is the ambient room temperature. In addition, it can be a function of the PS fan speed. However, in our PSs, it is accounted for by the fact that the fan speed is a linear function of $I$, which has been added as an argument of $P^{\mathrm{(diss)}}_{ijk}$. It is important to note that some APS power supplies had water cooling, whereas all APS-U power supplies will be air cooled. Hence, water cooling is not considered in our model.

During our experiments, it was determined that $T_\mathrm{room}$ cannot be assumed constant. Its temporal variations must be accounted for in the model to achieve sufficient prediction performance. However, one can represent the temperature increment in Eq.~\eqref{eq:finite_difference_heat_equation_0} as 
$\Delta T_{ijk} =  \Delta(T_{ijk} - T_\mathrm{room}) + \Delta T_\mathrm{room}$ and use the fact that in practice $\abs{\Delta T_\mathrm{room}}\ll\abs{\Delta(T_{ijk} - T_\mathrm{room})}$. Alternatively, $\abs{\dv{T_\mathrm{room}}{t}}\ll\abs{\pdv{T(x,y,z,t)}{t}}$. Hence, $\Delta T_\mathrm{room}$ can be neglected, and one can perform a variable change $(T_{ijk} - T_\mathrm{room})\longrightarrow T_{ijk}$, turning Eq.~\eqref{eq:finite_difference_heat_equation_0} into
\begin{multline}
  \begin{aligned}\label{eq:finite_difference_heat_equation}
  \Delta T_{ijk} = F_{ijk}(I,T_{ijk})
  &+ \widehat{\alpha}_{ijk} \; (T_{i+1\;j\;k} - T_{ijk}) + \widehat{\alpha}_{i-1\;j\;k} \; (T_{i-1\;j\;k} - T_{ijk})
  \\
  &+ \widehat{\beta}_{ijk} \; (T_{i\;j+1\;k} - T_{ijk}) +
  \widehat{\beta}_{i\;j-1\;k} \; (T_{i\;j-1\;k} - T_{ijk})
  \\
  &+ \widehat{\gamma}_{ijk} \; (T_{i\;j\;k+1} - T_{ijk}) +
  \widehat{\gamma}_{i\;j\;k-1} \; (T_{i\;j\;k-1} - T_{ijk}),
  \end{aligned}
\end{multline}
\noindent where now $T_{ijk}$ represents the difference between the temperature of cube $i,j,k$ and the ambient room temperature, $\widehat{\alpha}_{ijk}=\Delta t\; \alpha_{ijk} / C_{ijk}$, $\widehat{\beta}_{ijk}=\Delta t\; \beta_{ijk} / C_{ijk}$, $\widehat{\gamma}_{ijk}=\Delta t\; \gamma_{ijk} / C_{ijk}$, and
\begin{equation}
  F_{ijk}(I,T_{ijk}) = \Delta t \left[P^{\mathrm{(heat)}}_{ijk}(I) - P^{\mathrm{(diss)}}_{ijk}(I,T_{ijk})\right] / C_{ijk}.
\end{equation}

\begin{figure} 
	\centering
		\includegraphics[width=\textwidth]{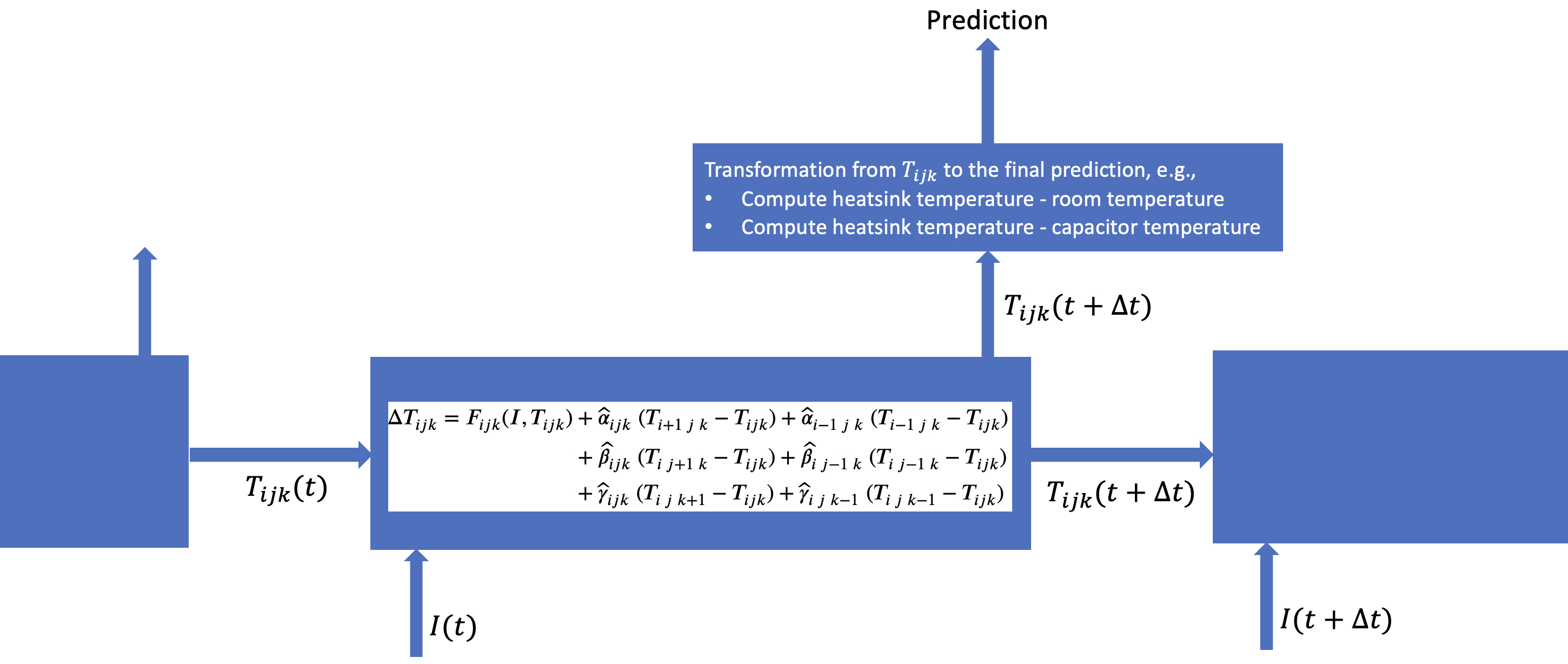}
	  \caption{Similarities between solving the heat equation using the finite differences method and a recurrent neural network.}\label{fig:heat_equation_as_rnn}
\end{figure}

The similarity between the recurrent relation of Eq.~\eqref{eq:finite_difference_heat_equation} and a recurrent neural network (RNN) becomes clear when the former is represented as a box diagram shown in Fig.~\ref{fig:heat_equation_as_rnn}. This similarity between RNNs and numerical methods for time-dependent partial differential equations (PDEs) has been pointed out before in other recent publications \cite{alt2023,zubov2021neuralpde,huang2022partial}.
The values $T_{ijk}$ correspond to the hidden state of an RNN cell, because typically we only 
know the temperatures at a few points in the PS, where temperature sensors are installed. For 
example, our PSs have sensors for the heatsink temperature, capacitor temperature, and damping 
resistor temperature. The transformation that converts the hidden state into a prediction (see 
the top box in Fig.~\ref{fig:heat_equation_as_rnn}) can be understood then as selecting, e.g., the 
heatsink temperature (minus room temperature) element from the hidden state $T_{ijk}$. If room 
temperature is unavailable (which is essential for adjusting the built-in sensor readings to account for variations in room temperature in order to enhance model performance), one can compute the differences between the available temperature   
readings instead, e.g., heatsink temperature minus capacitor temperature, damping resistor temperature minus capacitor temperature, etc.

\newcommand{\nunits}{n_u}
\subsection*{Long Short-Term Memory Model}

In this paper, we propose to replace the recurrent relation [see Eq.~\eqref{eq:finite_difference_heat_equation} and the bottom-center box in Fig.~\ref{fig:heat_equation_as_rnn}] by an LSTM cell \cite{lstm}, and to replace the transformation for the prediction (the top box in Fig.~\ref{fig:heat_equation_as_rnn}) by a fully connected layer of neurons. We chose LSTM, because it is the most commonly used RNN type for time-series forecasting \cite{KHALDI2023119140}. Custom-built architectures might perform better. However, our aim was to show that using a standard, pre-built RNN cell can still achieve strong results. Therefore, we use the Tensorflow \cite{tensorflow} implementation of an LSTM cell with hyperbolic tangent activation function and $\nunits$ units, where $\nunits$ is a variable hyperparameter. A fully connected hidden prediction layer (the top box in Fig.~\ref{fig:heat_equation_as_rnn}) is used with $\nunits$ units as well and a rectified unit activation function.
The output prediction layer does not use an activation function and may consist of either a single unit or multiple units, depending on the quantity of temperature variables predicted.
However, we focus on predicting exclusively the PS heatsink temperature for the sake of simplicity, while it still sufficiently demonstrates the method's performance.
Henceforth, this model will be referred to as the LSTM model.

Interestingly, when the PS board is monitored with an IR camera, all of the temperatures in the PS are known [in the two-dimensional approximation of Eq.~\eqref{eq:finite_difference_heat_equation}]. Hence, there is no need for a hidden state (of an RNN cell). Therefore, the (two-dimensional) function $F_{ij}$ and the coefficients $\alpha_{ij},\,\beta_{ij},\,\gamma_{ij}$ can be inferred more easily, without resorting to RNNs. This will be considered in Section~\ref{sec:ir_camera}.

\newcommand{\npoly}{n_p}
\subsection*{Effective Thermal Capacity Approximation Model}

Another way to eliminate the need for a hidden state is to approximate the PS by just a single cube 
in Eq.~\eqref{eq:finite_difference_heat_equation}, with the temperature of this cube being the 
temperature that the model is trying to predict, e.g., the PS heatsink temperature (minus room temperature).
The function $F$ can be approximated by a polynomial of finite degree $\npoly$. Then, for this single cube, Eq.~\eqref{eq:finite_difference_heat_equation} takes the form (with the convention $0^0=1$)
\begin{equation}\label{eq:etca}
  \Delta T = F(I, T) = \sum\limits_{n=0}^{\npoly}\sum\limits_{m=0}^{n}f_{m\;n-m}I^mT^{n-m}.
\end{equation}
\noindent where the coefficients of the polynomial $f_{m\;n-m}$ can be found using linear regression methods. This model approximates the thermal properties of the PS by a 
single effective thermal capacity and by a single effective temperature. Therefore, this model will 
be referred to as the Effective Thermal Capacity Approximation (ETCA). Despite its simplicity, this model's 
prediction performance is rather good. It will be used as the baseline model, when assessing the 
performance of the LSTM model below in this paper.

\section{Apparatus and Data Collection}\label{sec:psTestStand}

To assess the optimal architectures of the LSTM and ETCA models, and to evaluate their comparative performance, a specialized power supply test stand was constructed. This setup includes a power supply (PS) for a corrector magnet of APS-U storage ring, a PS controller, a resistive load for the PS, an ambient temperature sensor, and two relays for managing the cooling fans of the PS, complemented by a high-resolution infrared (IR) camera. A Raspberry Pi computer facilitated the (a) acquisition and network broadcast of IR camera imagery at a rate of approximately 9 frames per second, (b) collection and broadcast of the built-in PS parameter data and the ambient temperature readings, and (c) programmed control of the relays to operate the PS cooling fans. By intermittently deactivating the fans, we induced synthetic anomalies in the PS temperature to test the performance of the LSTM and ETCA anomaly detection techniques.

\begin{figure} 
	\centering
		\includegraphics[width=0.75\textwidth]{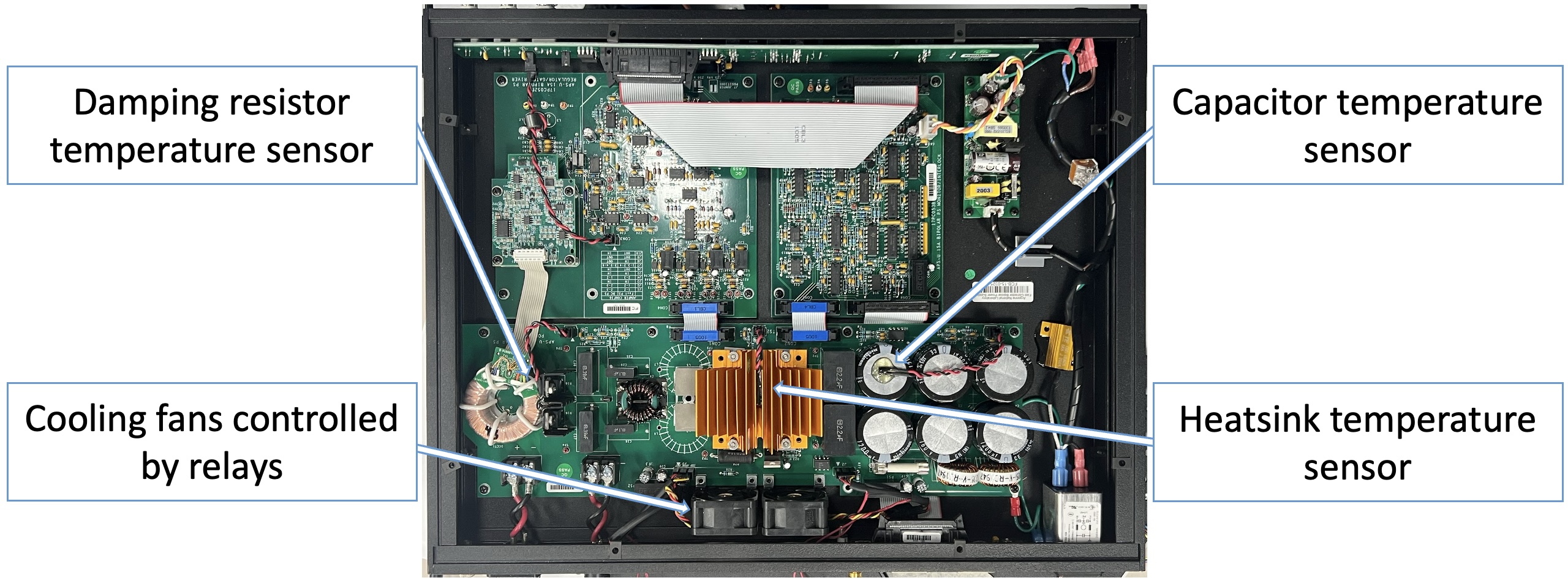}
	  \caption{APS-U storage ring corrector-magnet power supply.}\label{fig:PS_photo}
\end{figure}

The PS chassis constitutes a flat box with a detachable lid.
For our initial proof-of-concept experiments utilizing the high-resolution IR camera, the lid was removed to allow for direct observation of the internal components.
The IR camera, specifically a Seek-Thermal camera \cite{seek-thermal} with a resolution of $320\times 240$ pixels, was strategically positioned above the PS to ensure its field of view closely aligned with the imagery presented in Fig.\ref{fig:PS_photo}. 
We will 
briefly discuss the long-term vision for employing such more advanced PS temperature monitoring in Section~\ref{sec:ir_camera}. However, the main focus of this paper is 
the anomaly detection methods for the built-in temperature sensors, shown in 
Fig.~\ref{fig:PS_photo}, because they require installation of little to no additional equipment, which is important for scaling these methods to the thousands of APS-U power supplies.

\begin{figure} 
	\centering
		\includegraphics[width=0.9\textwidth]{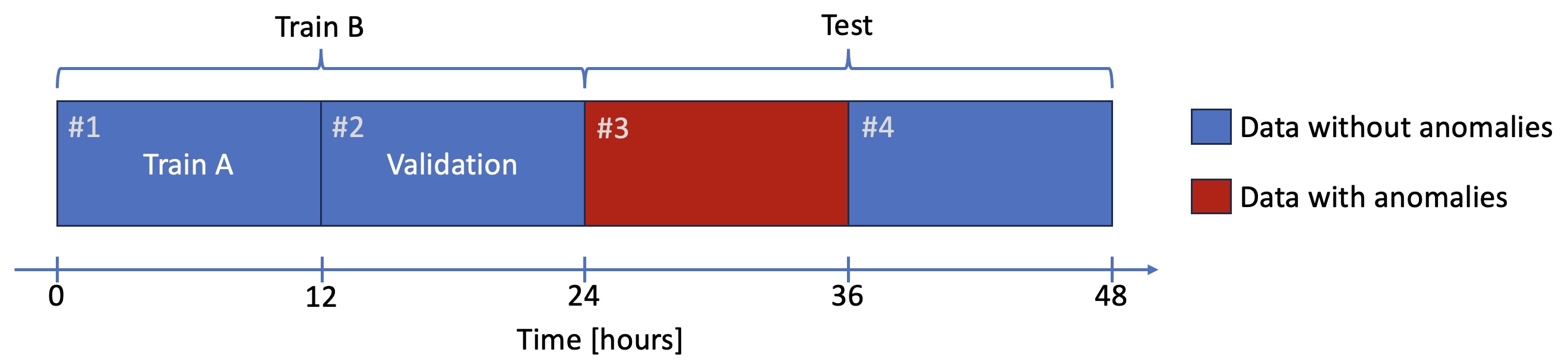}
	  \caption{Train-validation-test data split.}\label{fig:data_split}
\end{figure}
The structure of the considered dataset is outlined in Fig.~\ref{fig:data_split}. The 
overall length is 48 hours at a rate of one sample per second. This dataset can be divided 
into four 12-hour blocks serving different purposes. Each of the four blocks is 
subdivided into two 6-hour halves with two different types of PS current behavior. The first type of behavior is a series of random linear ramps, plateaus, and step changes. The second type is a random walk with reflection at \SI{0}{A} and \SI{15}{A}. The first data type was chosen because it is quite representative of normal PS operation. The second data type was added because it encourages the LSTM model to learn the fundamental principles of heat exchange and dissipation in the power supply, as opposed to merely memorizing how the PS temperature behaves during the linear ramps, plateaus, and step changes. We are restricting our considerations only to the positive polarity currents for the sake of simplicity. Generalizing the models to both positive and negative polarity currents is discussed in Appendix~\ref{sec:pos_and_neg_currents}.
\begin{figure}
	\centering
		\includegraphics[width=\textwidth]{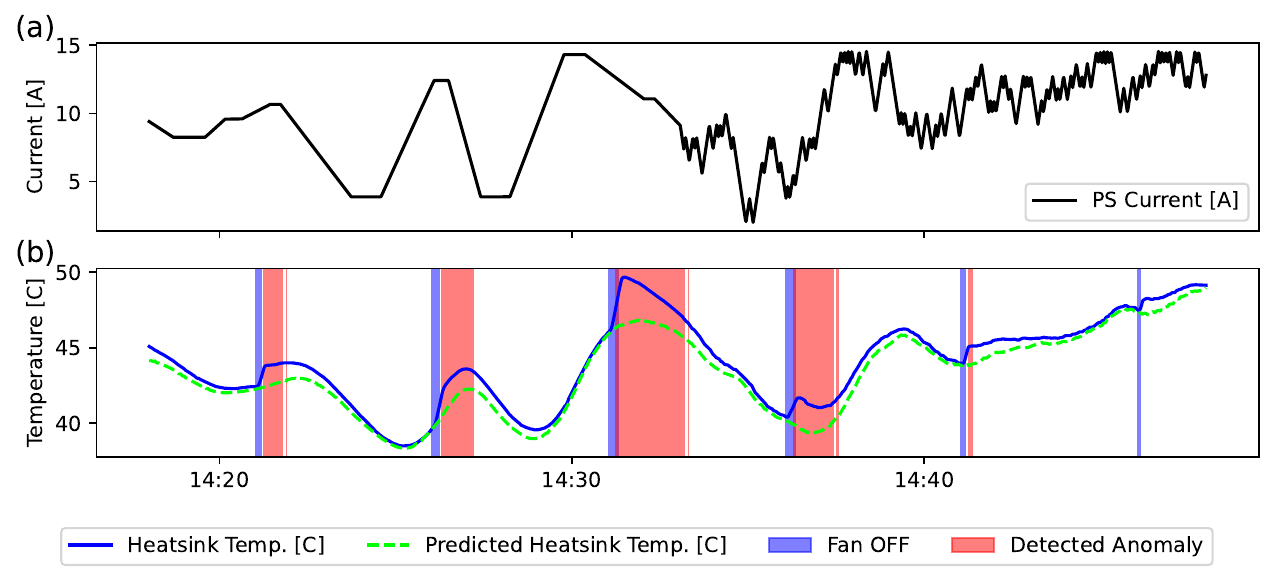}
	  \caption{Synthetic anomalies in the heatsink temperature created by temporarily turning off the cooling fan.}\label{fig:fan_off_anomalies}
\end{figure}
The first and the second types of PS current behavior are illustrated in the left and right halves of Fig.~\ref{fig:fan_off_anomalies}(a), respectively. The time interval shown in Fig.~\ref{fig:fan_off_anomalies}(a), corresponds to 30 minutes in the middle of Block \#3 from Fig.~\ref{fig:data_split}. The intentional temperature anomalies in Block \#3 were generated by turning off the PS cooling fans every 5 minutes for a brief interval of time. The duration of this interval was random and was drawn from a uniform distribution from 5 to 20 seconds. In Fig.~\ref{fig:fan_off_anomalies}(b) such intervals are represented by the blue shaded regions.

\newcommand{\HeatsinkTempRMSE}{\SI{0.16}{C}}
\newcommand{\HeatsinkTempEffRMSE}{\SI{0.11}{C}}
\newcommand{\HeatsinkTempEffCapRMSE}{\SI{0.08}{C}}
Before training any models, we conducted a small experiment to assess the repeatability of the PS heatsink temperature response to identical PS current waveforms.
We considered two 5-hour intervals of identical random PS current waveforms, separated by 12 hours, and the corresponding PS temperature responses.
The total range of change of the heatsink temperature was  from 33 to $\SI{50}{C}$. The Root Mean Square Error (RMSE) between the two 5-hour heatsink temperature responses was $\HeatsinkTempRMSE$. This is the ultimate precision any model can achieve, if it tries to predict the heatsink temperature directly, based solely on the PS current time-series. This repeatability characteristic can be improved, if one considers a new variable---the heatsink temperature minus the room temperature. In this case, the RMSE becomes $\HeatsinkTempEffRMSE$. However, it should be noted that APS-U power supplies do not have built-in sensors for the room temperature (intake air temperature). Installing them in all power supplies is possible. However, we believe that an overall better decision will be to train the model to predict the differences between the built-in temperature sensors instead (heatsink, capacitor, damping resistor). It will make it harder to interpret the results of such an anomaly detection method, because further investigation of the anomalous time interval by a human will be necessary to determine which exact sensor has an anomaly (or multiple). However, this small disadvantage is better than the cost of installation of intake air temperature sensors in thousands of power supplies, and the risk of damaging them during the installation. As an example, for the difference between the heatsink temperature and the capacitor temperature, the RMSE between the two 5-hour repeatability-test intervals was $\HeatsinkTempEffCapRMSE$---even slightly better than $\HeatsinkTempEffRMSE$, corresponding to the heatsink temperature minus the room temperature.
Nonetheless, in what follows we focus on predicting the heatsink temperature minus the room temperature, because of the better interpretability of the results, which improves clarity of the description of this method.

\section{Hyperparameter Optimization}
The proposed LSTM model has one hyperparameter---$\nunits$, which is the number of units in the LSTM cell and the number of units in the fully-connected output layer (see Section~\ref{sec:theory}). The ETCA model also has one hyperparameter---the degree of the polynomial in the approximation of the function $F(I, T)$ (see Section~\ref{sec:theory}). 

\begin{figure} 
	\centering
		\includegraphics[width=\textwidth]{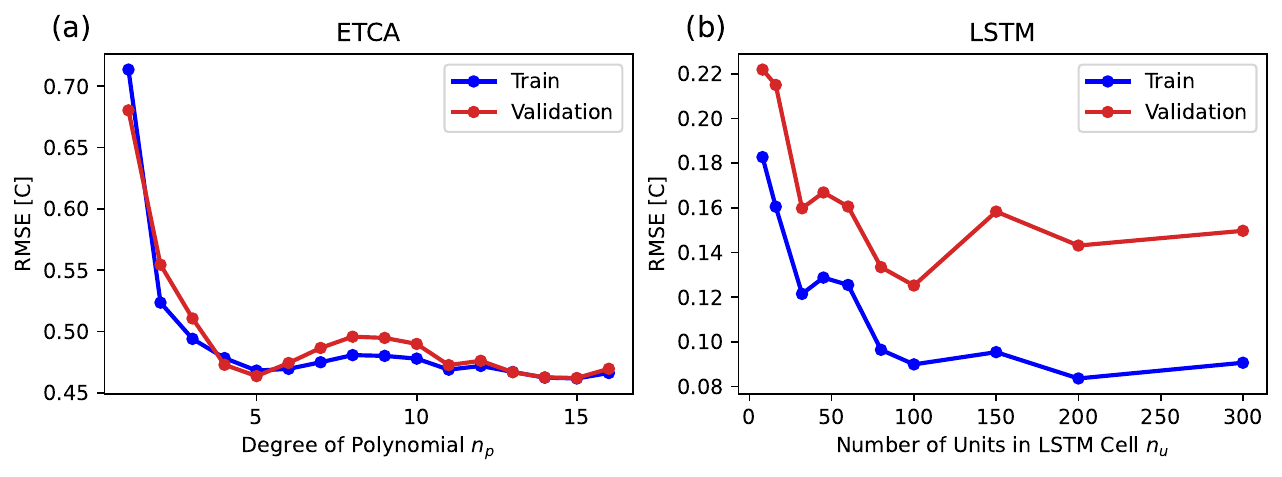}
	  \caption{Root Mean Square Error (RMSE) of the PS heatsink temperature at different hyperparameter values for the ETCA (a) and the LSTM (b) models.}\label{fig:hyperparameter_optimization}
\end{figure}

To perform hyperparameter optimization, Block \#1 (labeled as Train A) was used for training, and Block \#2 (labeled as Validation) was used for validation (see Fig.~\ref{fig:data_split}). The results of the hyperparameter exploration are shown in Fig.~\ref{fig:hyperparameter_optimization}. 
The RMSE presented in Fig.~\ref{fig:hyperparameter_optimization} is calculated when a trained model (ETCA or LSTM) is applied to an uninterrupted 12-hour long time-series of PS current values (Block \#1 for Train, and Block \#2 for Validation). The first 10 minutes of the time-series are ignored during the calculation of the RMSE (for both ETCA and LSTM models), because the LSTM model starts to perform well only after a sufficient amount of time, when the system forgets about the initial conditions, see Appendix~\ref{sec:init_irrel}.

For the linear regression of Eq.~\eqref{eq:etca} in the ETCA model, a Min-Max scaling was used, and the L1 regularization (lasso regression), with a small regularization parameter $\lambda = \SI{5e-6}{}$. Without regularization, at higher degrees of the polynomial approximation $\npoly$, overfitting occurs and some parameters of regression become too large, the model's predictions on the validation data become unstable and run away. As one can see in Fig.~\ref{fig:hyperparameter_optimization}(a), the RMSE on the validation data decreases significantly as $\npoly$ is increased from 1 to 5, and then it remains approximately the same. Since a model with fewer parameters is preferred, we choose 5 as the optimal polynomial degree for the ETCA model, $\npoly^\mathrm{(opt)} = 5$.

To obtain the results depicted in Fig.~\ref{fig:hyperparameter_optimization}(b), the LSTM model was trained for 300--450 epochs (depending on $\nunits$). During this training, we implemented a dynamic reduction in the learning rate upon encountering plateaus in the loss function. The training data was subdivided into 3000-second-long sequences, the first 600 seconds of which were ignored during the calculation of the loss function, by using zero weights. A
Min-Max scaler was used.
Figure~\ref{fig:hyperparameter_optimization}(b) illustrates an approximately monotonic decrease of the RMSE on the training data with increasing $\nunits$. Conversely, the RMSE on the validation data exhibits an optimal value at approximately $\nunits^\mathrm{(opt)}=100$, which will be used for subsequent analysis in this paper. This observed minimum highlights the trade-off inherent in selecting $\nunits$: while too few units compromise predictive accuracy, excessively many units lead to overfitting. Figure~\ref{fig:hyperparameter_optimization}(b) was created using the resources of the LCRC cluster at ANL, using the hyperparameter tuning library Ray Tune \cite{liaw2018tune}.

\section{Benchmarking LSTM's Anomaly Detection Performance Against the ETCA Model}\label{sec:benchmarking}
Upon identifying the optimal hyperparameter configurations for both the ETCA and LSTM models, these models underwent retraining using the consolidated training dataset, denoted as Train B in Fig.~\ref{fig:data_split}, which combines Block \#1 and Block \#2. These finalized models are tested on the Test dataset (see Fig.~\ref{fig:data_split}) composed of Block \#3 and Block \#4.
\newcommand{\LSTMtrainRMSE}{\SI{0.110}{C}}
\newcommand{\LSTMvalRMSE}{\SI{0.124}{C}}
\newcommand{\ETCAtrainRMSE}{\SI{0.497}{C}}
\newcommand{\ETCAvalRMSE}{\SI{0.453}{C}}
The RMSE performance of these models on the training data, Train B, is $\ETCAtrainRMSE$ and $\LSTMtrainRMSE$ for ETCA and LSTM models, respectively. The RMSE performance on the test Block \#4 (without anomalies) is $\ETCAvalRMSE$ and $\LSTMvalRMSE$ for ETCA and LSTM models, respectively. The LSTM model is considerably better than ETCA at predicting the PS heatsink temperature. Figure~\ref{fig:fan_off_anomalies}(b) presents the LSTM's predictions (dashed lime line) for the heatsink temperature (solid blue line). The blue shaded regions indicate the intervals where the cooling fans were turned off, and the red shaded regions represent the corresponding detected anomalies. The anomaly threshold (defined as the minimum deviation between measured and predicted heatsink temperatures to declare an anomaly) used in Fig.~\ref{fig:fan_off_anomalies}(b) was chosen to be relatively high, \SI{1.2}{C}, in order to include an example where an anomaly is not detected (see the right-most fan-off region).

\begin{figure}
	\centering
		\includegraphics[width=\textwidth]{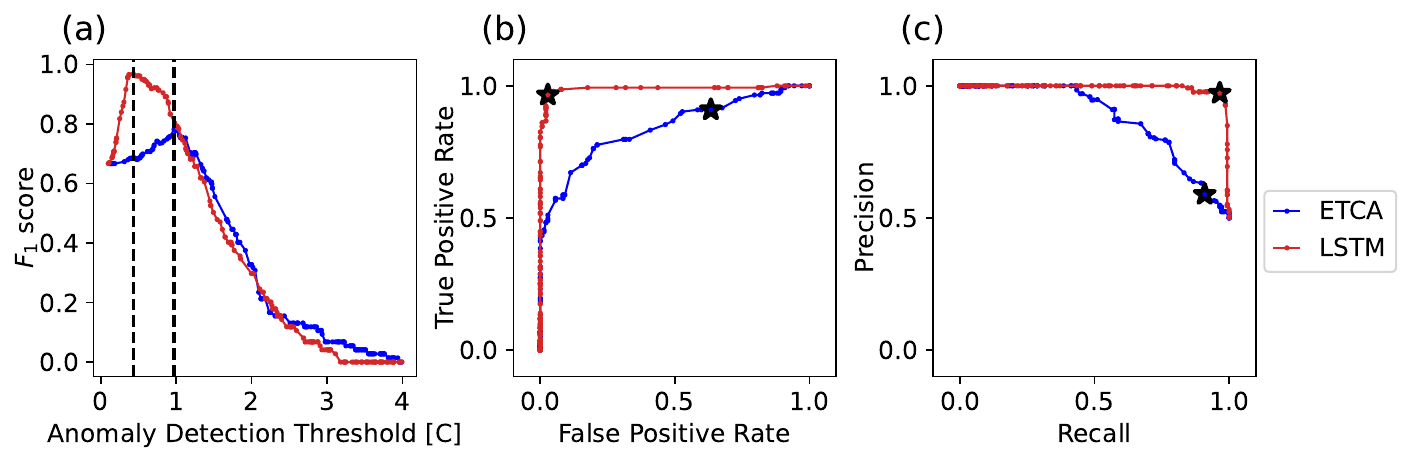}
	  \caption{Anomaly detection performance metrics for the ETCA and LSTM models at different anomaly detection threshold values. (a) $F_1$ score, (b) receiver operating characteristic (ROC) curve, (c) precision-recall curve.}\label{fig:F1_ROC_PRC}
\end{figure}

\newcommand{\ETCAthresOpt}{\SI{0.97}{C}}
\newcommand{\LSTMthresOpt}{\SI{0.43}{C}}
\newcommand{\ETCAfOneOpt}{\SI{0.78}{}}
\newcommand{\LSTMfOneOpt}{\SI{0.97}{}}
Further evaluation of the models' capability in anomaly detection was conducted as follows. The test dataset, denoted as Test in Fig.~\ref{fig:data_split} and spanning 24 hours, was segmented into 5-minute intervals. For Block \#3, the start of each interval aligned with the commencement of the fan-off periods. Subsequently, every 5-minute segment was treated as a separate instance for anomaly detection assessment, delineating four scenarios: (1) absence of an intended anomaly coupled with no anomaly detection, yielding a true negative; (2) presence of an intended anomaly without its detection, resulting in a false negative; (3) detection of an anomaly in the absence of an intended anomaly, leading to a false positive; and (4) detection of an anomaly when an intended anomaly was present, constituting a true positive. It is important to note that although the 5-minute intervals were evaluated independently, the ETCA and LSTM models made predictions for the PS heatsink temperature across the entire continuous 24-hour Test dataset.
Following the analysis of each 5-minute segment across various anomaly detection threshold values, we plotted three key performance metrics: the $F_1$ score, receiver operating characteristic (ROC) curve, and precision-recall curve, as depicted in Fig.~\ref{fig:F1_ROC_PRC}.
The maximum $F_1$ scores for the ETCA and LSTM models are $\ETCAfOneOpt$ and $\LSTMfOneOpt$, respectively. They are achieved at the anomaly detection threshold values $\ETCAthresOpt$ and $\LSTMthresOpt$, corresponding to the dashed vertical lines in Fig.~\ref{fig:F1_ROC_PRC}(a) and to the stars in Figs.~\ref{fig:F1_ROC_PRC}(b),(c).

\section{Future Outlook: Spatially-Resolved Anomaly Detection Using Infrared Cameras}\label{sec:ir_camera}
\begin{figure}
	\centering
		\includegraphics[width=\textwidth]{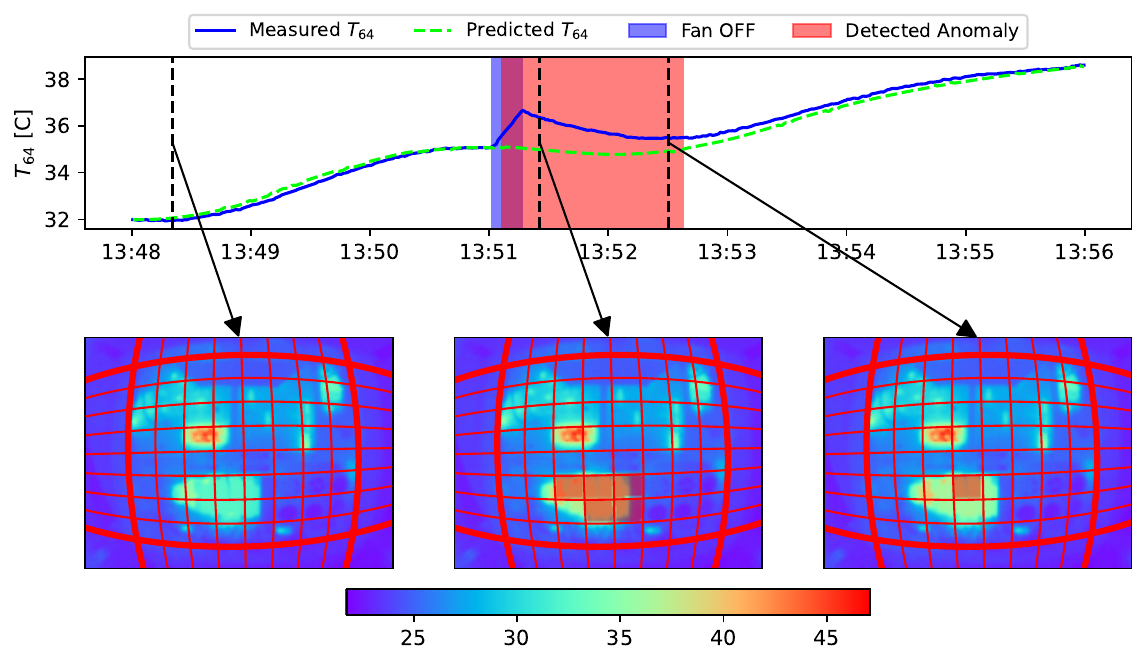}
	  \caption{Illustration of spatially resolved anomaly detection using the high-resolution infrared camera. $T_{64}$ represents the average temperature of the segment located 6th from the top and 4th from the left. The scale in the color bar is in degrees Celsius.}\label{fig:spatially_resolved_anomaly_detection}
\end{figure}

In the future, we are planning to develop more advanced temperature anomaly detection methods for power supplies employing infrared (IR) cameras, monitoring the interior temperature distributions. This technique can be very powerful even with very cheap and small low-resolution IR cameras, e.g., the $32\times24$ MLX90640 infrared cameras \cite{mlxThermalCamera}, which can be implemented in thousands of APS-U power supplies at a relatively low cost and minimal changes to the power supply design. However, in this paper, in Fig.~\ref{fig:spatially_resolved_anomaly_detection}, we present a proof-of-concept example of application of such a technique with a more expensive high-resolution $320\times240$ IR camera \cite{seek-thermal}.
In this example, the IR camera overlooks the power supply. The boundary formed by the wide red lines in the IR images in Fig.~\ref{fig:spatially_resolved_anomaly_detection} closely corresponds to the contour of the power supply. The orientation is the same as in Fig.~\ref{fig:PS_photo}. The red horizontal and vertical lines in Fig.~\ref{fig:PS_photo} were obtained as projections of equidistant meridians (of an auxiliary sphere) onto the flat image. This helps account for the barrel distortion effect of the IR camera. Average temperatures were calculated in all of the segments, formed by the red lines inside the power supply. Further, these averages (with the room temperature subtracted) were used as $T_{ij}$ values in the two-dimensional version of Eq.~\eqref{eq:finite_difference_heat_equation},
\begin{equation}\label{eq:finite_difference_heat_equation_2d}
	\Delta T_{ij} = F_{ij}(I,T_{ij})
	+ \widehat{\alpha}_{ij} \; (T_{i+1\;j} - T_{ij}) + \widehat{\alpha}_{i-1\;j} \; (T_{i-1\;j} - T_{ij})
	+ \widehat{\beta}_{ij} \; (T_{i\;j+1} - T_{ij}) +
	\widehat{\beta}_{i\;j-1} \; (T_{i\;j-1} - T_{ij}).
\end{equation}
We approximated the functions $F_{ij}(I,T_{ij})$ by polynomials as in Eq.~\eqref{eq:etca}, namely,
\begin{equation}\label{eq:ir_F_polynomial}
	F_{ij}(I, T) = \sum\limits_{n=0}^{\npoly}\sum\limits_{m=0}^{n}f^{(ij)}_{m\;n-m}I^mT^{n-m},
  \end{equation}
where $\npoly=3$ was used. Given this choice of $F_{ij}(I, T)$, Eq~\eqref{eq:finite_difference_heat_equation_2d} describes a linear regression problem with respect to $f^{(ij)}_{m\;n-m}$, $\widehat{\alpha}_{ij}$, $\widehat{\beta}_{ij}$. We used the ridge regression with a small L2 regularization parameter $\lambda=10^{-3}$, applied exclusively to $\widehat{\alpha}_{ij}$ and $\widehat{\beta}_{ij}$; no data scaling was performed. The ridge regression was chosen over the lasso regression, because it was much faster, while still sufficiently constraining the $\widehat{\alpha}_{ij}$ and $\widehat{\beta}_{ij}$ coefficients. After training, the model was applied to the test data. An example of an anomaly detected in Block \#3 of Fig.~\ref{fig:data_split} is presented in Fig.~\ref{fig:spatially_resolved_anomaly_detection}. An anomaly was declared (independently in each segment), when the measured average temperature in the segment deviated from the prediction by more than $\SI{0.5}{C}$. The top plot in Fig.~\ref{fig:spatially_resolved_anomaly_detection} shows the measured and predicted average temperature of one of the segments (6th from the top and 4th from the left) as a function of time. The three panels in the bottom represent three snapshots of the power supply interior at the three moments in time, indicated by the vertical black dashed lines in the top panel. The segments with anomalies are denoted by red shading. The color bar in the bottom represents the scale in degrees Celsius. 

One could argue that the heat exchange terms (with $\widehat{\alpha}_{ij}$ and $\widehat{\beta}_{ij}$) are unnecessary in Eq.~\eqref{eq:finite_difference_heat_equation_2d}, and that the segments can be considered independently with individual ETCA models as in Eq.~\eqref{eq:etca}. We considered this case and calculated the average (over all segments) RMSE for Block \#4 from Fig.~\ref{fig:data_split}. It was $\SI{0.125}{C}$. Whereas for the linear model with the heat exchange terms [Eq.~\eqref{eq:finite_difference_heat_equation_2d}] it was $\SI{0.118}{C}$. Therefore, including the heat exchange terms results in an improvement in the model prediction performance, although not a very significant one.

\section{Discussion and Conclusions}\label{sec:discussion_and_conclusions}

Figure~\ref{fig:F1_ROC_PRC} illustrates that the LSTM model is considerably better at detecting anomalies than the ETCA model. Consequently, the LSTM model emerges as the preferred method for anomaly detection in the thousands of APS-U magnet power supplies. However, it should be mentioned that we have carried out similar tests with another power supply from a different section of the APS accelerator complex, namely, a corrector-magnet power supply from the linear accelerator of the APS's injector complex. In this case, the performance of both the LSTM and ETCA models was comparable, albeit the LSTM model still performed slightly better. The ETCA model has some advantages, namely, the training speed and the guarantee of finding the global minimum, since Lasso  regression is almost instant in this case and is convex. Whereas training the LSTM model takes 10--30 minutes on a typical workstation and requires careful choice of the learning rate in the process.

Implementation of the LSTM models in the thousands of APS-U power supplies will begin when APS-U enters routine operation after all critical commissioning milestones and goals have been achieved.
In this paper, we employed an external room temperature sensor, whose reading was subtracted from the PS's built-in temperature sensors. This improved model performance, since these quantities are less sensitive to room temperature variations, see Section~\ref{sec:psTestStand}.
However, this additional room temperature sensor can be avoided if one considers a model that predicts the differences between heatsink, capacitor, and damping resistor temperatures instead.
While this approach may make anomaly interpretation somewhat more challenging and may not effectively detect anomalies with uniform deviations across all built-in temperature sensors, it offers a considerable benefit in terms of scalability. Specifically, it will allow for easy expansion across the thousands of APS-U storage-ring power supplies without requiring the installation of extra equipment.

Section~\ref{sec:ir_camera} demonstrated the potential of using infrared cameras for monitoring the interior temperature distributions of the power supplies. This approach not only yields more precise details regarding the location of identified anomalies but also enhances the detection of anomalies that might elude the built-in temperature sensors. Infrared cameras, data storage, and processing power for machine learning have become significantly cheaper in the last few decades, making it feasible to implement such diagnostics in the thousands of APS-U power supplies in the future.
In Section~\ref{sec:ir_camera}, the demonstration of our proof-of-concept utilized a high-resolution IR camera, which can be substituted with one or more cost-effective, lower-resolution IR cameras. Our future efforts will focus on fine-tuning the choice and positioning of the camera(s) within the power supply and improving the machine learning algorithms for anomaly detection using IR camera data. We intend to develop a final design that balances performance and cost-effectiveness suitable for implementation across all APS-U power supplies.
These results will be reported in future publications.

\section*{Acknowledgments}

The authors extend their sincere gratitude to Andrew Kreps and Ju Wang from ANL for their invaluable technical assistance in constructing the power supply test stand. We also wish to express our appreciation to Nikita Kuklev from ANL and to Jonathan Edelen and Kathryn Wolfinger from Radiasoft for their insightful contributions through numerous productive discussions. We gratefully acknowledge the computing resources
provided on Bebop and Swing, high-performance computing clusters operated by the Laboratory Computing Resource Center
at Argonne National Laboratory.
Argonne National Laboratory is a U.S. Department of Energy laboratory managed by UChicago Argonne, LLC. 
This research used resources of the Advanced Photon Source, a U.S. Department of Energy (DOE) Office of Science user facility at Argonne National Laboratory and is based on research supported by the U.S. DOE Office of Science-Basic Energy Sciences, under Contract No. DE-AC02-06CH11357.




\appendix
\section{Asymptotic Irrelevance of Initial Conditions}\label{sec:init_irrel}
Consider the ETCA model with the following approximation for the dissipated power $P^{\mathrm{(diss)}}=\kappa\;T$, keeping in mind that $T$ represents the difference between the PS temperature and the room temperature. Then,
\begin{equation}\label{eq:simple_diss_model}
  C_\mathrm{eff}\dv{T}{t} = P^\mathrm{(heat)}(t) - \kappa\;T.
\end{equation}
This equation has an exact solution using Laplace transforms,
\newcommand{\tp}{t^\prime}
\begin{equation}\label{eq:example_of_exact_solution}
  T(t) = T(0) e^{-t/\tau}
+ \frac{1}{C_\mathrm{eff}}\int\limits_0^t e^{-(t-\tp)/\tau} P^\mathrm{(heat)}(\tp) \dd{\tp},
\end{equation}
\noindent where $\tau=C_\mathrm{eff}/\kappa$, and $T(0)$ is the initial PS temperature at time $t=0$. Equation~\eqref{eq:example_of_exact_solution} shows that the system forgets about the initial conditions after a sufficient amount of time; and the characteristic time for this effect is $\tau$. The heat dissipation model of Eq.~\eqref{eq:simple_diss_model} provides a very clear illustration of the asymptotic irrelevance of initial conditions due to the exponential decay of the first term in Eq.~\eqref{eq:example_of_exact_solution}. However, this property goes beyond this simple model and it applies to the real PSs considered in this paper, for which the characteristic time $\tau$ is of the order of a few minutes. This fact allows us to use random initializers for the hidden states of the LSTMs considered in this paper. In this case the prediction performance is poor for the first few minutes (5--10), but it reaches optimal levels after that.

\section{ETCA and LSTM Models in Case of Positive and Negative PS Currents}\label{sec:pos_and_neg_currents}
In the main portion of the paper, the case of unipolar current is considered for the sake of clarity. In practice, the power supplies we considered were bipolar. It was noticed that the temperature response to positive and negative polarity currents can be slightly different. In principle, this can be accounted for by Eq.~\eqref{eq:etca}, which can also be re-written as (with the convention $0^0=1$)
\begin{equation}\label{eq:etca_appendix}
  \Delta T = \sum\limits_{n=0}^{\npoly}\sum\limits_{m=0}^{n}f_{m\;n-m}I^mT^{n-m}.
\end{equation}
\noindent  However, it may require an expansion to a relatively large polynomial degree $\npoly$, and the total number of regression coefficients $f_{n\;n-m}$ (for the ETCA model) scales as $(\npoly+1)\cdot(\npoly+2) / 2$\;. A better approach is to allow different values for the expansion coefficients for positive and negative polarity currents,
\newcommand{\Fp}{f^{\scaleto{(+)}{5pt}}}
\newcommand{\Ip}{I_{\scaleto{(+)}{5pt}}}
\newcommand{\Fm}{f^{\scaleto{(-)}{5pt}}}
\newcommand{\Iminus}{I_{\scaleto{(-)}{5pt}}}
\begin{equation}\label{eq:etca_bipolar}
  \Delta T = \sum\limits_{n=0}^{\npoly}\sum\limits_{m=0}^{n}\Fp_{m\;n-m} \Ip^m T^{n-m}
              + 
             \sum\limits_{n=0}^{\npoly}\sum\limits_{m=0}^{n}\Fm_{m\;n-m} \Iminus^m T^{n-m},
\end{equation}
\noindent where $\Ip = H(I)\cdot I$ and $\Iminus = \left(H(I) - 1\right)\cdot I$, and $H(I)$ is the Heaviside step function. For the right-hand side of Eq.~\eqref{eq:etca_bipolar} to be a continuous function of $I$ at $I=0$ and at any $T$, one must require $\Fp_{0\;q} = \Fm_{0\;q}$, $q=0,1,\dots,\npoly$\;. Therefore, the total number of independent regression coefficients is $(\npoly+1)\cdot(\npoly+2) / 2 - (\npoly+1)$\;. More importantly, sufficient accuracy can be achieved with a smaller polynomial degree $\npoly$ [relative to Eq.~\eqref{eq:etca_appendix}].

\printcredits

\bibliographystyle{unsrtnat}

\bibliography{main}



\end{document}